\documentstyle[11pt,AATS,epsf]{article}	% DO NOT TOUCH

\begin{document}	% DO NOT TOUCH 

%-----------------------------------------------------------------------
%       Paper Title 
%-----------------------------------------------------------------------
% Enter the title of the paper. Please use mixed cases.
\title{Dust to Dust: Evidence for Planet Formation?} 

%-----------------------------------------------------------------------
%	Author(s) names and affiliation
%-----------------------------------------------------------------------
% Enter the authors followed by their affiliations.  The \author and
% \affil commands may appear multiple times as necessary. List each
% author by giving the first name or initials first followed by the last
% name.  Authors with the same affiliations should be grouped together,
% with the names separated by commas (no &, no 'and').
\author{Glenn Schneider, Dean C. Hines, Murray D. Silverstone}
\affil{Steward Observatory, University of Arizona}
\author{Alycia J. Weinberger, Eric E. Becklin}
\affil{Division of Astronomy \& Astrophysics, UCLA}
\author{Bradford A. Smith}
\affil{Institute for Astronomy, University of Hawaii}

%-----------------------------------------------------------------------
%	Contact information
%-----------------------------------------------------------------------
% Please include a comment (a line which starts with a %) with the name
% and email address of a contact person in case the editors experience
% difficulties with your manuscript.
% First name, Last Name, email address
% Glenn, Schneider, gschneider@as.arizona.edu

%-----------------------------------------------------------------------
%                              Abstract
%-----------------------------------------------------------------------
% Type abstract in the space below, between the "\begin{abstract}" and
% "\end{abstract}" lines. No blank line after the "\begin{abstract}"
% line! 
\begin{abstract}

We discuss the properties of several circumstellar debris disk systems 
imaged with the Hubble Space Telescope's Near Infrared Camera and 
Multi-Object Spectrometer in a survey of young stars with known far-IR 
excesses. These dusty disks around young ($\sim$ 5--8 Myr) unembedded stars 
exhibit morphological anisotropies and other characteristics which are 
suggestive of recent or on-going planet formation. We consider evidence 
for the evolution of populations of collisionally produced disk grains in 
light of the significant presence of remnant primordial gas in the 
optically thick disk of the classical T-Tauri star TW Hya; the 
dust-dominated and Kuiper-belt like circumstellar ring about 
the young main-sequence star HR 4796A; and a possible ``intermediate'' 
case, the complex disk around the Herbig AeBe star HD 141569A. Only a
small number of debris disks have been imaged thus far in scattered light. Yet all 
show structures that may be indicative of dust reprocessing, possibly as 
a result of planet formation, and speak to the contemporary competing 
scenarios of disk/planet evolution and formation time scales. 
\end{abstract}

%-----------------------------------------------------------------------
%                             Main Body
%-----------------------------------------------------------------------
% After the abstract comes the main body of the manuscript. Use \section
% to label various sections. You may also use \subsection. Please
% use mixed cases in the sections' titles. Sections and subsections will
% be numbered automatically. 

\section{Introduction}

Warm dust around young stars has been inferred from thermal infrared excesses
since IRAS, though until recently the expected ``cold'' dust component
had been imaged in circumstellar scattered light only about $\beta$  Pictoris.
To begin to understand both the evolution and dynamics of
circumstellar disks, and the nascent planetary systems which they may harbor,
a multi-wavelength attack is required.  Mid-IR observations can reveal
and trace emission from warm dust.  Colder disk components can be detected,
and on larger spatial scales mapped, in the sub-millimeter.  Resolved
images of dusty disks, at near-IR and optical wavelengths, 
provide direct information on the spatial distribution of the disk grains.
Until recently, however, such observations have been extremely difficult, given the
very high disk-to-star contrast ratios, but when successful provide measurements
of radial and azimuthal asymmetries in the dust brightness distributions and
scattering properties (colors, phase functions, etc.).  The presence of
rings, gaps, clumps, warps and central holes seen in
scattered light images of circumstellar disks may implicate the existence 
of embedded or co-orbital perturbers.

Studied extensively for fourteen years since first imaged by
Smith \& Terrile (1984), the disk around the Vega-like star $\beta$  Pictoris served 
as the archetype and sole example of a dusty debris system seen in scattered light.  
Though now thought to be $\sim$ 20 Myr old, throughout this
period its age remained uncertain (and controversial) and was estimated
from a few hundred Myr to about 10 Myr. The $\beta$ Pictoris disk was
found to possess at least five asymmetries attributed speculatively to 
dynamical interactions with unseen planetary bodies (Kalas \& Jewitt 1995).

Space-based near-IR and optical coronagraphic imaging
with the Hubble Space Telescope second generation instruments NICMOS
and STIS has provided a powerful new tool for studying the 
circumstellar environments of nearby stars.
Though only a very small number of spatially resolved images of
dusty disk systems currently exist, these systems exhibit a diversity in disk
sizes, morphologies, and properties (e.g., Figure 1). 

The presence and morphologies of disk asymmetries and the derived properties
and spatial distributions of the constituent grains may help to better 
constrain the ages and evolutionary status of the possible fledging 
extra-solar planetary systems. The presumed time scales for disk evolution 
will be tested and refined with an accumulation of observations of disk systems
such as these.

%%%%%%%%%%%%%%%%%%%FIGURE%%%%%%%%%%%%%%%
\begin{figure}[h]
\includegraphics{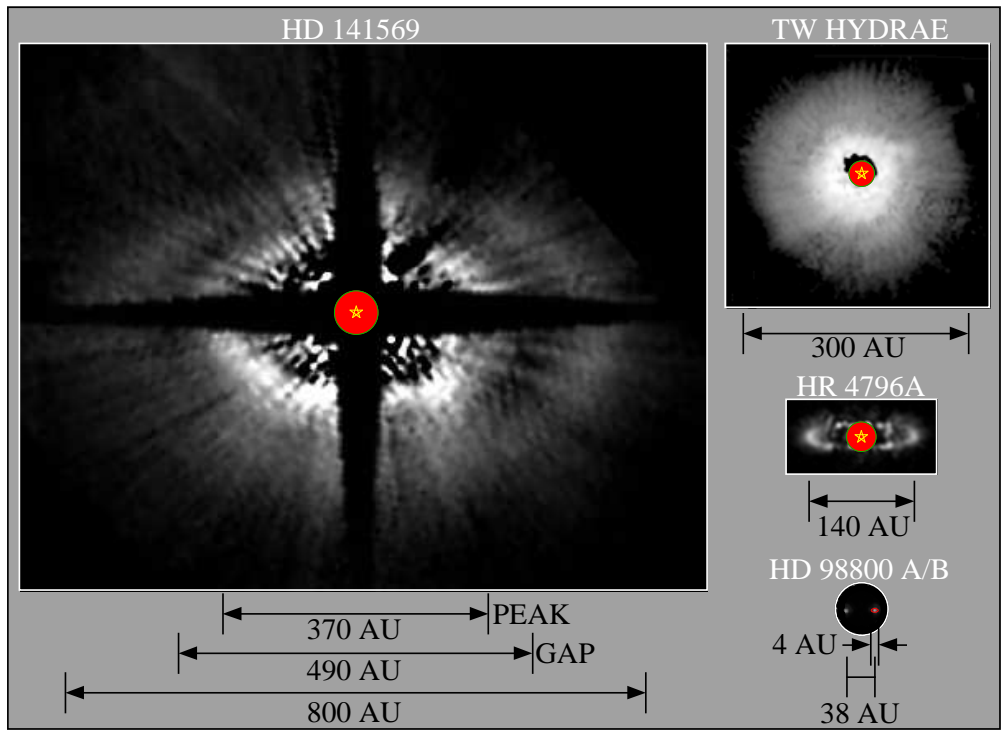}
\vspace{9.4truecm}
\caption[]{Comparative sizes and morphologies of dust/debris systems 
around $\sim$  5-8 Myr stars imaged by NICMOS.}
\end{figure}
%%%%%%%%%%%%%%%%%%%%%%%%%%%%%%%%%%%%%%%%

\section{Disk Ages and Evolution}

It is commonly conjectured that following the early stages of 
protostellar collapse, as the rocky cores of giant planets form (in 
$\sim$ 0.1--1 Myr), primordial dust in the circumstellar environment is 
both  dominated (in the ratio of $\sim$ 100:1) by and locked to gas in 
the disk. Current theories of circumstellar disk  evolution suggest a 
presumed epoch of planet-building, on the order of 1 Myr following the
protostellar collapse, via the formation and agglomerative growth of 
embryonic bodies. Gaseous atmospheres will then accrete onto the giant 
planet cores on time scales of a few to about 10 Myr attendant with an 
expected significant decline in the gas-to-dust ratios in the 
remnant protostellar environments. During these times primordial dust 
(i.e., ISM-like grains) would be cleared from ``typical'' systems 
on shorter time scales by radiation pressure ($\sim$ $10^{4}$ yr), 
and by Poynting-Robertson drag ($\sim$ 1 Myr). In these critical evolutionary 
phases of newly-formed (or still forming) extra-solar planetary systems,  
the circumstellar environments become dominated by a second-generation 
dust population containing larger grains replenished through the
collisional erosion of planetesimals, and perhaps, by cometary infall. 
As the circumstellar regions become optically thin (and the central 
stars become largely unembedded) the likely-evolving population of dusty 
debris  at these early epochs become more readily observable in 
scattered light. This scenario suggests a morphological evolutionary 
sequence which could be modified by the perturbing influence of co-spatial 
bodies, and which may be explored (and validated) by high contrast imaging.

The presumption of evolution, however, requires a knowledge of the 
(relative) ages of disk systems, but such knowledge is not 
necessarily secure.  Determining the ages of PMS and ZAMS stars depends 
strongly upon transforming observable quantities for 
placement on HR diagrams, and finding their isochronal ages with respect 
to their birthlines based upon theoretical evolutionary tracks.  The 
fidelity (or lack thereof) of the stellar evolutionary models, and 
measurement errors in the observables (distances, luminosities, spectral 
temperatures, etc.) both contribute to uncertainties in derived ages 
(see Figure 2), which can be significant, particularly for earlier 
(Vega-like) stars (e.g., $\beta$  Pic, HR 4796A, HD 141569A). The 
situation is improved if late spectral type coeval companions can be 
found, as is the case for HR 4796A (Jura et al. 1993) and HD 141569A 
(Weinberger et al. 2000), which are discussed in this paper.  
Three of the four disk systems we discuss (HR 4796A, TW Hya, and HD 
98000A/B) are likely members of the TW Hya association (Webb et al. 1999),
the nearest site of recent star formation to the earth.  
Together with HD 141569A, the small sample of young disk systems we 
discuss here are of similar ages ($\sim$ 5--8 Myrs), yet 
have very different properties.

%%%%%%%%%%%%%%%%%%%FIGURE%%%%%%%%%%%%%%%
\begin{figure}[h]
\includegraphics{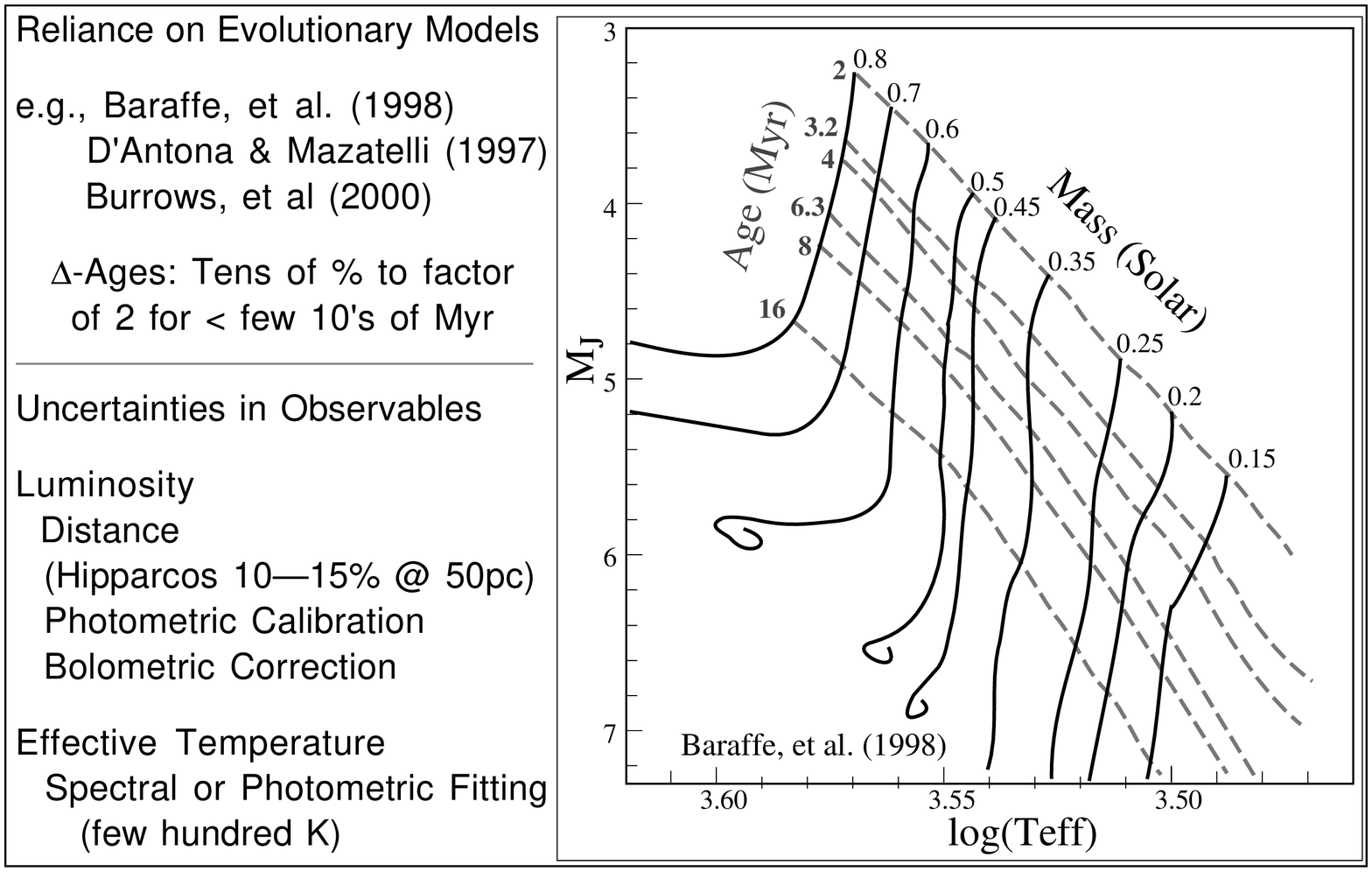}
\vspace{7.9 truecm}
\caption[]{Stellar (and hence, disk) age determinations are dependent 
upon uncertainties in both stellar models and measurements.}
\end{figure}
%%%%%%%%%%%%%%%%%%%%%%%%%%%%%%%%%%%%%%%%

\section{Four $\sim$  5--8 Myr Disk Systems}

Dusty disks were spatially resolved and 
imaged around three young ($<$ 10 Myr old) stars.  These disks show
radial and hemispheric brightness anisotropies and 
complex morphologies, both possibly indicative of dynamical interactions 
with unseen planetary mass companions. From these and other 
observations we describe and compare of the properties of these dusty 
debris systems:

a) TW Hya (a K7Ve T-Tau ``old'' PMS star) with a pole-on circularly 
nearly-symmetric disk, a radial break in its surface density of 
scattering particles,  and possibly a radially and azimuthally confined 
arc-like depression.

b) HD~141569A (a Herbig Ae/Be star, $\sim$ 5 Myr) with a 400 AU radius 
inclined disk with a 40 AU wide gap at 250~AU.

c) HR~4796A (a A0V star, $\sim$ 8 Myr) with a 70 AU radius ring less 
than 14 AU wide and unequal ansal flux densities.

Additionally, our non-detection of scattered light and high precision 
photometry of a fourth system of similar age, HD98800~A/B, coupled with 
mid-IR imaging and longer wavelength flux density measurments, 
greatly constrain a likely model for the debris about the B component. 

\subsection{TW Hydrae (TWA 1)}

TW Hya is an isolated classical T-Tauri star (Rucinski \& Krautter 
1983), exhibiting characteristic
H$\alpha$ and UV excesses and an IRAS far-IR
excess with $L_{\rm disk}/L_{\star}$  $\sim$  0.3.  Both sub-millimeter 
continuum (Weintraub et al. 1989) and CO (Zuckerman et al. 1995)
emission have been observed.  With a Hipparcos determined
distance of 56 $\pm$ 7 pc, and estimated age of 6 My, TW
Hya is the archetypal member of the young stellar association 
which bears its name.  These characteristics
made TW Hya a prime candidate for scattered light disk
imaging.

We found TW Hya to harbor an optically thick face-on disk (r $\sim$ 
190 AU) seen in NICMOS F110W and F160W coronagraphic images, also imaged by
Krist et al. (2000) with WFPC-2.  Additionally, we recently 
observed TW Hya coronagraphically in the optical with STIS, and find the 
azimuthally averaged surface brightness profile is globally fit very 
well at all wavelengths with an r$^{-2.6}$ power law.  The disk exhibits 
essentially gray scattering, implicating a characteristic particle size 
from the NIR colors of at least 2$\mu$m, probably evolved
from a primordial ISM-grain population.  Mid-IR spectroscopy
of this disk, obtained at Keck by Weinberger et al. (2001),
shows a broad $\sim$ 10$\mu$m emission from amorphous silicates.  
Silicates of a few microns in size can
explain the color, thermal extent, and shape of the mid-IR
spectrum and further implicate grain growth from the
original ISM population.

The brightness profile
is explained by an outwardly flared disk with a central
hole. Areal scattering profiles in the NIR bands reveal a break in the 
surface density of scatterers at R $\sim$ 105 AU, which may be 
indicative of sculpting of the disk grain distribution.
At the same radius, an arc-like depression confined
to about 90\deg\ in azimuthal extent is seen in
both the higher spatial resolution STIS and WFPC-2 images
(see Figure 3).  This feature might arise from shadowing of
the grains due to a discontinuity in the z-height distribution
of the flared disk, or to relative deficit of scattering
particles at that location. Either might arise from the
gravitational effects of an embedded perturber.

%%%%%%%%%%%%%%%%%%%FIGURE%%%%%%%%%%%%%%%
\begin{figure}[h]
\includegraphics{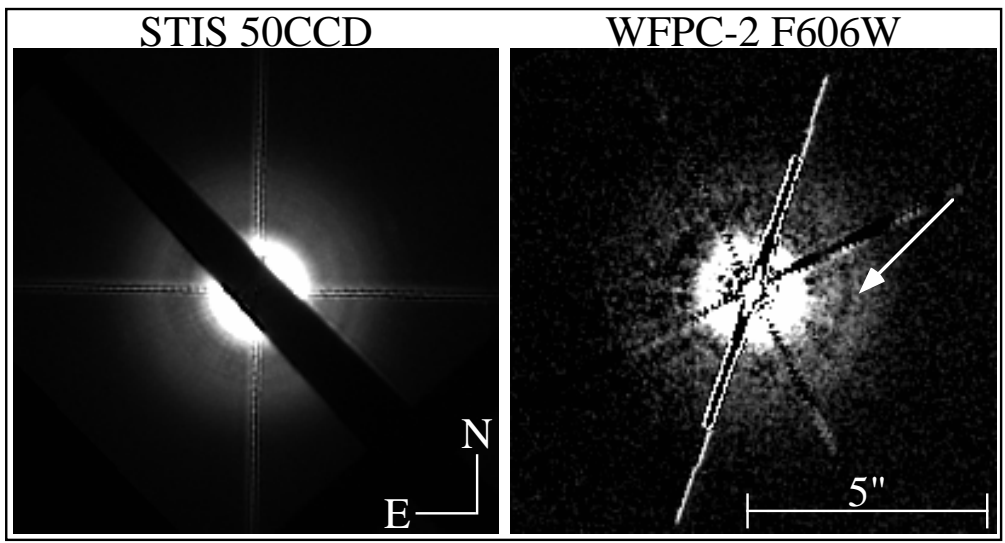}
\vspace{7.2truecm}
\caption[]{TW Hya circumstellar disk. Left - STIS (0.59 $\mu$m, FWHM 
0.45$\mu$m). Right - WFPC-2 (0.60 $\mu$m, FWHM 0.20$\mu$m) at the same scale 
and orientation.  Arrow indicates location of the dark arc-like feature 
(see text), also seen in the STIS image.}
\end{figure}
%%%%%%%%%%%%%%%%%%%%%%%%%%%%%%%%%%%%%%%%

\subsection{HD 141569A }

The Herbig Ae/Be star HD 141569 (B9V, H=6.89, d = 99 $\pm$ 8pc, 
2.3 M$_\odot$) was found to possess extended mid-IR emission
to a radius of $\sim$ 1\farcs2 at 12.5--20.8 $\mu$m 
Marsh et al. (2001).  NICMOS 1.1 $\mu$m 
coronagraphic observations revealed a scattered light disk to a radius 
of at least 400 AU, exhibiting a complex morphology including a 40 AU 
wide gap in the surface brightness profile at a radius of 250 AU
(Weinberger et al. 1999).
The disk, with a total 1.1 $\mu$m flux density of 8$\pm$2 mJy beyond 
0\farcs6 (peak surface brightness 0.3 mJy arcsec$^{-2}$ at 185 AU) is 
inclined to our line-of-site by 51\deg $\pm$ 3\deg.  Augereau et al. 
(1999) saw a similar morphology
in lower-resolution 1.6 $\mu$m observations, also obtained with NICMOS.  
No significant amount of scattered light was detected closer than r $\sim$  
1\farcs2, so the regions of warmer dust probed in the mid-IR, and the 
outer (colder) regions imaged in scattered light are mutually exclusive.

The intrinsic scattering function of the disk results in a brightness 
anisotropy in the ratio 1.5$\pm$0.2, with the brighter side in the 
direction of forward scattering. The region of the gap may be partially 
cleared of material by an unseen co-orbital planetary companion. If so, 
the width/radius ratio of the gap implies a planetary mass of $\sim$ 1.3 
Jupiters. This is consistent with a $<$ 3 Jupiter mass point-source
detection limit at this radius, where we also estimate the albedo to be 
0.35$\pm$0.05. HD 141569A is the brightest member of an $\sim$  6.5:1 
hierarchical triple system ($\Delta$A($\overline{BC}$) = 8\farcs3, 
$\Delta$$\overline{BC}$ = 1\farcs3), where the presumed coeval 
M-dwarf dynamical companions were 
used by Weinberger et al. (2000) to improve the age estimate for the 
disk system, now $\sim$ 5 Myr.  The companions probably influence 
the dynamics of the circumstellar disk.

\subsection{HR 4796A (TWA 11A)}

On 15 March 1998 NICMOS obtained the first scattered light image of a 
circumstellar debris disk since the discovery of the $\beta$  Pictoris 
disk.  With an age of 8 $\pm$ 2 Myr (Stauffer 1995), a spectral type of 
A0V, and possessing an M-dwarf companion, HR 4796A is similar to HD 
141569A, yet the structure of its disk is very different. The dust in 
the disk about HR 4796A was found to be confined in a narrow ring 70 AU 
in radius, as had been suggested from mid-IR imaging by Koerner et al. 
(1998) and Jayawardhana et al. (1998), and $<$ 14 AU in width (see Figure 
1).  Earlier, Jura (1991) inferred the presence of large amount of 
circumstellar dust from IRAS excess and estimated $L_{\rm disk}/L_{\star} 
\sim  5\times 10^{-3}$ (twice that of $\beta$ Pictoris). Jura et al. 
(1995) noted their earlier 110K estimate of dust temperature indicated a 
lack of material at $<$ 40 AU, and inferred grain sizes $>$ 3$\mu$m at 
40 $<$ r $<$ 200 AU given the time scales for disk-clearing.  Schneider et 
al. (1999) reported on the morphology, geometry and photometry of the 
ring-like disk from well-resolved NICMOS coronagraphic images at 1.1 and 
1.6$\mu$m from which, unlike TW Hya, the disk grains appeared to be red 
({\it J}(F110W)-H(F160W) = 0.6 $\pm$ 0.2). Augereau et al. (1999) 
successfully reproduced the observed properties of the disk from all of 
the then-available observations in a two-component model with a) cold 
amorphous (Si and H$_{2}$0 ice) grains $>$ 10$\mu$m in size (cut-off in size 
by radiation pressure), with porosity $\sim$0.6, peaking at 70 AU and b) 
hot dust at $\sim$ 9 AU of ``comet-like'' composition (crystalline Si and 
H$_{2}$0), porosity $\sim$ 0.97.  They noted that collisionally evolved 
gains, with bodies as large as a few meters, were required in their 
model which also gave rise to a minimum mass of a few earth masses with 
gas:dust $<$ 1. This is consistent with subsequent sub-millimeter 
observations by Greaves et al. (2000) wherein they estimated the total 
mass in gas as 1--7 earth masses.  This is also consistent with 
planetesimal accretion calculations by Kenyon \& Wood (1999) in which 
they find planet formation at 70 AU is possible in 10 Myr in an initial 
10--20 minimum mass solar nebula where dust production is then confined 
to a ring with $\Delta a$ = 7--15 AU.  Possible evidence for one (or 
more) unseen planets exists in an $\sim$ 10--15\% brightness asymmetry in 
the NE and SW ansae of the ring seen both in the NICMOS images and by 
Telesco et al. (1999) in 18.2$\mu$m OSCIR images, suggesting a 
pericentric offset possibly due to a gravitational perturber.  While HR 
4796B, an M-dwarf companion at a projected distance of 500 AU, may 
serve to truncate the outer radius of the disk, the narrowness of the 
ring might implicate co-orbital companions confining the dust through a 
process akin to the shepherding of ring particles in the Saturnian 
system.

\subsection{HD 98800A/B (TWA4 A/B)}

HD 98800, historically classified as a binary comprised of two similar K 
dwarfs (currently separated by 0\farcs8), was found by IRAS to contain 
one of the brightest far-IR excesses in the sky.  Now a recognized 
member of the TW Hya Association with a Hipparcos determined distance of 
46.7 $\pm$ 6pc, the two PMS components are themselves spectroscopic 
binaries with periods (Aa+Ab) = 262 days, (Ba+Bb) = 315 days (Torres 
et al 1995) and separations of $\sim$ 1 AU.  Gehrz et al. (1999) showed 
the debris is centered on the B component from 4.7 and 9.8$\mu$m 
observations, and 20\% of the luminosity of B is emitted in a 164 $\pm$ 
5K SED from the mid-IR to the sub-millimeter.  The SED is fit very well 
by a single temperature black-body, indicating that the grains 
co-exist in a very limited radial zone from the central stars. 
High-precision NICMOS photometry straddling peak of stellar SEDs by 
Low, Hines \& Schneider (1999)
find T$_{eff}$(A) = 3831 $\pm$ 55K,  T$_{eff}$(B) = 3459 $\pm$ 37K, and no 
detectable 0.9--1.9m$\mu$ excess.  From this they suggest the scattered:total light 
from B is $<$ 6\% implying an albedo of $<$ 0.3 for the debris system with 
an inner radius of 4.5 AU, subtending $\sim$ 20\% of the sky seen 
from the B component.  Koerner et al. (1999) confirmed the Ba+Bb 
circumbinary disk, and that the B components are the source of the large IR excess 
upon which a silicate feature is imposed.  From mid-IR imaging they 
suggest a disk with properties similar to Low et al. (1999).  Unblended
optical (0.5--1.0 $\mu$m) spectra of the A and B components we recently
obtained with STIS indicate that the B component closely resembles an M0V
star, so it slightly later in spectral type than previously thought 
(but similar in this regard to TW Hya).

The inferred geometry and properties of the HR 98800B disk bears a 
resemblance to the Zodiacal bands in our own solar system, and may be 
similar to the debris system around our Sun as it appeared a few million 
years after formation.  The multiplicty of the system undoubtedly 
complicates the dynamics, and hence the temporal stability and evolution 
of the grains.  The small size of the B-component circumbinary debris 
system may be causally related to interactions of the grains with the 
multiple components in the system.

\section{Summary}
The four dusty disks considered here exhibit significant variations in 
sizes, morphologies, and grain properties, despite their similar ages,
 and differ as well from $\beta$  Pictoris 
(of very similar spectral type to HR 4796A and HD 141569A).  The desire 
to construct a ``morphological evolutionary sequence'' for dusty debris 
disks is physically complicated by the dispersions in stellar spectral 
types (and hence masses), compositions and densities of the parent 
molecular clouds, and disk interactions with stellar and sub-stellar 
companions.  In addition, uncertainties in age determinations by as much as 
a factor of about two from 
observable diagnostics and theoretical models, further muddy the waters.  
The sample of such disks observed to date is very small.  Obviously, 
many more observations are needed to advance our detailed understanding 
of disk/planet formation mechanisms and time scales.

%-----------------------------------------------------------------------
%                             References
%-----------------------------------------------------------------------
% List your references below between the \begin{references} and
% \end{references} commands. Each reference should begin with a
% \reference command. Observe the following order when listing
% bibliographical information for each reference:  author name(s),
% publication year, journal name, volume, and page number for
% articles.  See the User's Guide for a list of macros to represent
% journals. 

% The acknowledgments section is optional.
\acknowledgments
We thank the other members of the NICMOS IDT for their many contributions
to our nearby stars programs, and to Eliot Malumuth, Phil Plait, and
Sally Heap for their valuable help with our STIS observations. This work is 
supported in part by NASA grant NAG 5-3042 and based on 
observations with the NASA/ESA Hubble Space Telescope.
\end{document}